\documentclass[sn-mathphys]{sn-jnl}

\jyear{2021}%

\theoremstyle{thmstyleone}%
\theoremstyle{thmstyletwo}%

\theoremstyle{thmstylethree}%

\usepackage{mhchem}
\usepackage{subcaption}
\begin{document}

\title[Article Title]{Single and two-phase fluid droplet breakup in impulsively generated high-speed flow}


\author*[1]{\fnm{James} \sur{Leung}}\email{jleung3@lsu.edu}

\author[1]{\fnm{Mohana} \sur{Gurunadhan}}\email{mgurun1@lsu.edu}
\equalcont{These authors contributed equally to this work.}

\author[1]{\fnm{Shyam} \sur{Menon}}\email{smenon@lsu.edu}
\equalcont{These authors contributed equally to this work.}

\affil*[1]{\orgdiv{Mechanical and Industrial Engineering Department}, \orgname{Louisiana State University}, \orgaddress{\city{Baton Rouge}, \postcode{70803}, \state{LA}, \country{USA}}}




\abstract{Aerobreakup of fluid droplets under the influence of impulsively generated high-speed gas flow using an open-ended shock tube is studied using experiments and numerical simulations. Breakup of mm-sized droplets at high Weber number was analyzed for water and two-phase nanofluids consisting of dispersions of \ce{Al2O3} and \ce{TiO2} nanoparticles in water with high loading of 20 and 40 weight\% respectively. Droplet breakup is visualized using high-speed imaging in the experimental setup, where an open ended shock tube generates impulsive high-speed flow impinging on a droplet held stationary using an acoustic levitator. Axisymmetric simulations using the Volume of Fluid technique are conducted to capture the gas dynamics of the flowfield and droplet deformation at the initial stages. Fluid droplets are subject to a transient flow field generated by the open ended shock tube, characterized by a propagating incident shock wave, a recirculating vortex ring, and standing shock cells. Droplet breakup for all fluids proceeds through an initial flattening of the droplet followed by generation of a liquid sheet at the periphery in the presence of a curved, detached shock front at the leading edge. The breakup appears to follow a sheet stripping process whereby stretched ligaments undergo secondary atomization through viscous shear. Mist generated in the wake of the droplet appears to expand laterally due to the unconstrained expansion of the high-speed gas jet. The breakup morphology of droplets for all fluids appears consistent with previous observations using conventional shock tubes. Lateral deformation of the coherent droplet mass is observed to be higher for nanofluids as compared to water. This is attributed to higher viscosity and Ohnesorge number of nanofluid droplets, which results in delayed breakup and increased lateral stretching. When plotted as a function of non-dimensionalized time, the same effects are also attributed to generate highest non-dimensional velocities for the \ce{TiO2} nanofluid, followed by \ce{Al2O3} nanofluid, and water, which mirrors the order of viscosity and Ohnesorge number for the three fluids. An area of spread, which can be interpreted as a measure of dispersion, plotted as a function of non-dimensionalized time also shows the highest value for the \ce{TiO2} nanofluid, followed by \ce{Al2O3} nanofluid, and water. Overall, current results indicate that droplet breakup for two-phase fluids appear to be similar to those for single-phase fluids with effectively higher viscosity. Further, an open ended shock tube proves to be an effective tool to study droplet aerobreakup, with some differences observed in the droplet wake due to the unconfined expansion of gas flow.}




\maketitle

\section{Introduction}\label{sec1}

The interaction of high speed gas with liquid droplets leading to their subsequent aerodynamic breakup (aerobreakup) has been studied by a number of researchers using experimental and computational techniques~\cite{nicholls1969aerodynamic, wierzba1988experimental,hirahara1992experimental,joseph1999breakup,theofanous2008physics,theofanous2011aerobreakup,theofanous2012physics,meng2015numerical,meng2018numerical}. The motivation for these studies includes droplet interactions with flows generated by aircraft or turbine blades moving at high speeds, secondary atomization processes in sprays, and liquid propellent rocket motor combustion instabilities~\cite{nicholls1969aerodynamic}. Recognizing that control of aerobreakup-driven droplet atomization can only be achieved after a complete fundamental understanding of the phenomena, much of the previous research has focused on issues such as: characterizing droplet breakup regimes, measuring breakup times, and evaluating drag coefficients. Droplet aerobreakup has been traditionally classified into different regimes on the basis of the Weber number ($We=\rho D_0 u_{\infty}^2/\gamma$). Theofanous and Li~\cite{theofanous2008physics} re-classified previous results primarily into two regimes: Rayleigh-Taylor piercing (RTP) and shear induced entrainment (SIE) with SIE being the terminal regime for $We>1000$. RTP is characterized by bag, bag-stamen, and multi-bag breakup modes driven by the Rayleigh-Taylor instability (RTI), while SIE is characterized by sheet thinning/stripping modes driven by the Kelvin-Helmholtz instability (KHI). Droplet breakup time has been evaluated in previous work~\cite{gel1973singularities,hsiang1992near,nicholls1969aerodynamic,reinecke1970study} and attempts have been made to correlate data using a non-dimensional time~\cite{pilch1987use,simpkins1972water,nicholls1969aerodynamic}, $t^*=t \frac{u_g}{d_0} \sqrt{\frac{\rho_g}{\rho_l}}$, with $t^*=0$ defined as the instant when an incident shock associated with the high speed flow first reaches the droplet surface, $\rho_g$ and $u_g$ being density and velocity of gas behind the shock front, $\rho_l$ being liquid density, and $d_0$ the initial droplet diameter. Although breakup times have varied across a wide range in previous work, a few have observed $t_{br}^* \approx 1$ or 1--2~\cite{theofanous2008physics,theofanous2011aerobreakup}, particularly when the breakup regime exceeds the threshold for SIE.

The brief discussion regarding droplet breakup regime and droplet breakup time is pertinent to the present work, which considers two-phase fluids, consisting of a liquid with a dispersion of metal nanoparticles. The use of metal oxide nanoparticles to increase the energy content of a liquid fuel has been investigated by several researchers~\cite{law2012fuel,lee1992burning,kong2015combustion}, with application to internal combustion~\cite{mehta2014nanofuels}, gas turbine~\cite{kannaiyan2017effects}, and pulse detonation engines~\cite{palaszewski2006metallized}. Accordingly, the primary motivation for the present work is to study the aerobreakup of two-phase liquid droplets using experiments and simulations.
A secondary motivation is to consider the use of an open-ended shock tube to conduct droplet aerobreakup studies. Experimental studies of droplet aerobreakup have been primarily conducted in closed shock tubes. 
An open ended shock tube~\cite{haselbacher2007open,yu1996shock} is a variation of the conventional shock tube, where the driven section is open to the atmosphere. Open ended shock tubes with various exit area profiles have been used to study blast waves in tunnels and gun discharges~\cite{kim2003weak,phan1984effect}. The shock tube being open to the ambient allows for a relatively simpler design and easier implementation of imaging diagnostics. While the open ended shock tube does produce a rapidly dissipating shock wave expanding into the ambient, earlier work has shown that droplet breakup is primarily driven by the high-speed gas flow generated in the post-shock region~\cite{meng2015numerical,hanson1963shock,nicholls1969aerodynamic} and the normal shock wave itself has little effect on droplet breakup. 
However, the high speed flow from an open ended shock tube is characterized by standing shocks, a jet boundary, and a multidimensional flow field~\cite{haselbacher2007open}. To ensure a comprehensive understanding of droplet breakup driven by impulsive flows generated by an open ended shock tube, the underlying gas dynamics and its impact on the breakup process needs to be understood. 

The layout of the paper is as follows. The experimental setup utilizing an open ended shock tube is described along with the approach used to position droplets using an acoustic levitator. Diagnostics employed for fluid flow and droplet breakup visualization including high-speed imaging and Schlieren are described, and details of the test fluids and operating conditions are listed. Details of the simulation approach are provided and numerical methods used in the solution process are discussed. Results obtained from experiments and simulations for the flowfield and gas dynamics are discussed first. This is followed by results discussing the phenomenology of droplet breakup as well as dynamics of the droplet mass as it undergoes breakup. Finally, conclusions that can be drawn from the results are summarized.

\section{Experimental Approach}\label{sec2}
\subsection{Setup}\label{subsec21}
\begin{figure}[h!]
\centering
\includegraphics[scale=0.55]{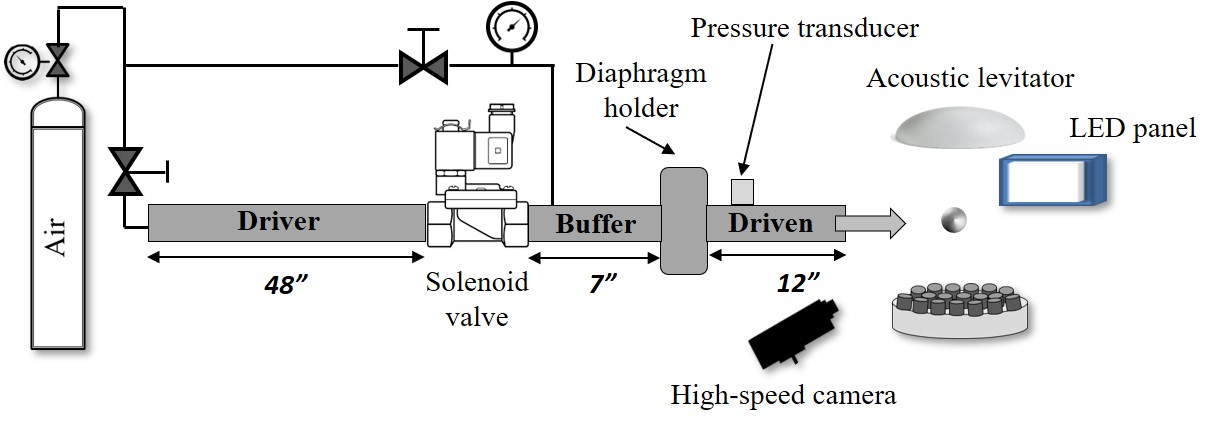}
\caption{Schematic of experimental setup shown with the direct imaging setup.}\label{fig:Setup}
\end{figure}

Figure~\ref{fig:Setup} shows a schematic of the experimental setup. The open ended shock tube comprises of three main sections. A driver section is 48" long and is supplied air from a compressed air cylinder with a pressure regulator. The driver is separated from an intermediate buffer section by an electrically actuated solenoid valve. The buffer section is 7" long. A diaphragm holder assembly with a polycarbonate diaphragm separates the buffer section from the driven section. The driven section is 12" long. All three sections are constructed using 1" diameter pipe. The rationale for this design is discussed next.

Given the limitations on camera recording time at the desired frame rates (20 kfps), it is needed to synchronize the impulsive flow generation with the start of camera recording. For this purpose, an electrically activated solenoid valve was initially chosen to separate the driver and driven sections with no buffer section in the tube. However, tests showed that flow restrictions in the solenoid valves resulted in reduction in flow throughput leading to weaker shocks. This issue has been previously reported in the context of solenoid valve use in shock tubes~\cite{lynch2016note}. Shock strength is found to be directly correlated to valve opening time and flow throughput. To improve the flow throughput while retaining the solenoid valve, it was decided to implement a buffer region, itself separated from the driven section using a polycarbonate diaphragm. The buffer region pressure can be independently controlled in the setup shown in Fig.~\ref{fig:Setup}. Solenoid valve opening is used as a means to trigger the camera recording, while the use of the buffer zone allows high throughput flow of air into the driven section once the diaphragm ruptures. The driver pressure is set to 120 psig for all experiments. For an identical driver pressure of 120 psig, a $\sim$25\% increase in pressure was noted at the dynamic pressure sensor for the case with the diaphragm, as compared to the case with solenoid valve only. The buffer pressure choice was investigated and found to have no significant impact on the characteristics of the generated shock wave. Accordingly, the buffer pressure was set to 60 psig for all tests.  

A polycarbonate diaphragm, 0.005" in thickness, is used at the buffer-driven section interface. To conduct a test, the driver and buffer sections are filled with air to the required pressures. Once the solenoid is activated, and the diaphragm ruptures, a shock wave and accompanying high-speed flow is generated. The flow is directed towards a droplet kept levitated using an acoustic levitator. The 3D printed single-axis levitator~\cite{marzo2017tinylev} consists of an array of ultrasonic emitters installed on mirroring top and bottom curved surfaces. The system has a total of 72 transducers whose location and inclination are selected such that they create pressure nodes where a droplet can be trapped and held in a levitated position. The transducers are controlled using input generated by an Arduino Nano and operate at a frequency of 40 kHz. Droplets of the fluid to be tested are levitated in the levitator unit prior to generating the shock wave. Best results are obtained by dispensing the fluid using a syringe fitted with a needle tip. 
The droplets tend to be elliptical in shape, due to the action of acoustic waves, which transform the spherical droplet into an ellipsoid~\cite{watanabe2018contactless}. Typical aspect ratio for a levitated ellipsoidal droplet ranges from 1.4--1.6.
 
\subsection{Diagnostics}\label{subsec22}

Droplet breakup upon being exposed to the impulsively generated high speed air flow is visualized using two forms of imaging. A high-speed camera (Photron SA-3) is used with a diffuse LED light source to capture images of the droplet breakup process. Imaging was carried out at 20,000 fps with a resolution of 512x128 pixels. Frame rate and resolution were selected to achieve the highest number of images of the droplet breakup process before the liquid mass leaves the field of view of the camera. A resolution of 5--6 pixel/mm is obtained using a 150 mm focal length lens on the camera. The camera is triggered manually and is synchronized with the opening of the solenoid valve. Schlieren imaging was also conducted using the same high-speed camera and a traditional Z-type Schlieren setup. The setup consists of two 6” mirrors of 60” focal length with an LED providing the light source. A set of plano-convex lenses mounted in line with the LED provides the optics needed for the light source. The camera is focused on the droplets levitated in the test section. Pressure measurements in the driven section of the tube were acquired using a dynamic pressure transducer (PCB Model 113B22) along with a signal conditioning module. 

\subsection{Test fluids and operating conditions}\label{subsec23}

Tests were conducted using three fluids: water, aluminum oxide (\ce{Al2O3}) nanofluid, and titanium dioxide (\ce{TiO2}) nanofluid. \ce{Al2O3} nanofluid consisted of spherical \ce{Al2O3} nanoparticles, 30 nm in diameter, 20 wt\% dispersed in water. \ce{TiO2} nanofluid consisted of spherical \ce{TiO2} nanoparticles, 30 nm in diameter, 40 wt\% dispersed in water.

\begin{table}[h]
\caption {Fluid properties and operating conditions.}
\centering
\small
\tabcolsep=0.11cm
\begin{tabular}{| p{3.35cm} | p{1.75cm} | p{2.7cm} | p{2.7cm} |} 
\hline
   \textbf{Property} &\textbf{Water} & \textbf{\ce{Al2O3} nanofluid} & \textbf{\ce{TiO2} nanofluid}\\ \hline
  Density [kg/m3] & 1000 & 1590 & 2292\\\hline
  Viscosity [Pa.s] & 0.0008 & 0.022 & 0.18\\\hline
  Surface tension [mN/m] & 70.32 & 62 & 60.5\\\hline
  We & 14,000 & 15,870 & 16,400\\ \hline
  Oh & 0.0021 & 0.0486 & 0.3448\\ \hline
  Re & \multicolumn{3}{c|}{262,000} \\ \hline
       \hline  
\end{tabular}
\label{Tab:FluidProps} 
\end{table}
Table~\ref{Tab:FluidProps} summarizes the fluid properties and key non-dimensional variables relevant to the study. Gas properties ($\rho_g$=3.5 $kg/m^3$, $\mu_g$=1e-5 Pa.s) and velocity ($u_g$=375 $m/s$) required for calculation of We and Re are evaluated using results from flow simulations. The values used correspond to gas properties at at the exit of the shock tube. An average droplet size of 2 mm is used for the calculations shown in Table~\ref{Tab:FluidProps}. Surface tension for all fluids was measured at ambient conditions using a tensiometer (Attension Theta by Biolin Scientiﬁc). Viscosity of nanofluids was extrapolated from measurements presented in previous work~\cite{shar16,chand10,turg09,sekh15}. We number values indicate that droplet breakup occurs in the Shear Induced Entrainment (SIE) breakup regime as per Theofanous (We$>10^3$)~\cite{theofanous2008physics}. 
A steady increase in Oh is noted going from water to \ce{Al2O3} nanofluid to \ce{TiO2} nanofluid, primarily caused by increase in fluid viscosity. Oh, which represents the ratio of viscous to inertial and capillary forces influences the critical We ($\mathrm{We}_{cr}$), above which breakup would occur for a liquid droplet exposed to air flow of constant or increasing velocity. As Oh increases, $\mathrm{We}_{cr}$ also increases~\cite{pilch1987use}. However, the operating We for all cases in this work are significantly higher than the $\mathrm{We}_{cr}$ ($\mathcal{O}(10)$) for the corresponding Oh, implying that aerobreakup of the droplet would indeed take place under the given conditions.

\section{Simulation approach}\label{sec3}
A two-step approach is used to simulate the droplet breakup process driven by the high-speed flow generated by the open-ended shock tube in a computationally efficient manner. A 2D axisymmetric single phase simulation is used to model the gas dynamics occurring within the shock tube and capture the gas expansion at open end. Transient variation of pressure and velocity, at the open end of shock tube, is imposed as the inlet boundary condition for a 2D axisymmetric two-phase simulation of the droplet breakup process, with phase transport and liquid-gas interface modeled using the Volume-Of-Fluid (VoF) technique. The two step approach significantly reduces the number of mesh points and computational effort otherwise needed to model the integrated gas-dynamic and droplet breakup processes, considering the disparate length scales and time scales involved in the problem. All simulations were conducted in ANSYS-Fluent. The two phase simulations were limited to water for primarily two physical reasons. One, nanofluid property measurements in literature are limited to cases with a maximum volume fraction of nanoparticles of $\sim$5\%~\cite{shar16,chand10,turg09}. Extrapolation of these properties to the case of nanofluids used in the present work with significantly higher loading may not be warranted. Two, there is some evidence in literature suggesting non-Newtonian behavior for nanofluids depending on applied shear rate, concentration, and nanoparticle size~\cite{shar16,nadoo18,chen07}. Given high shear rates in the present work and potential non-Newtonian behavior, simulations for nanofluid droplets could be significantly affected through the use of an inaccurate constitutive relationship. 

\subsection{Simulation domain and boundary conditions}\label{subsec31}

\begin{figure}[h!]
\centering
\begin{subfigure}{.85\textwidth}
  \centering
  \includegraphics[width=1\linewidth]{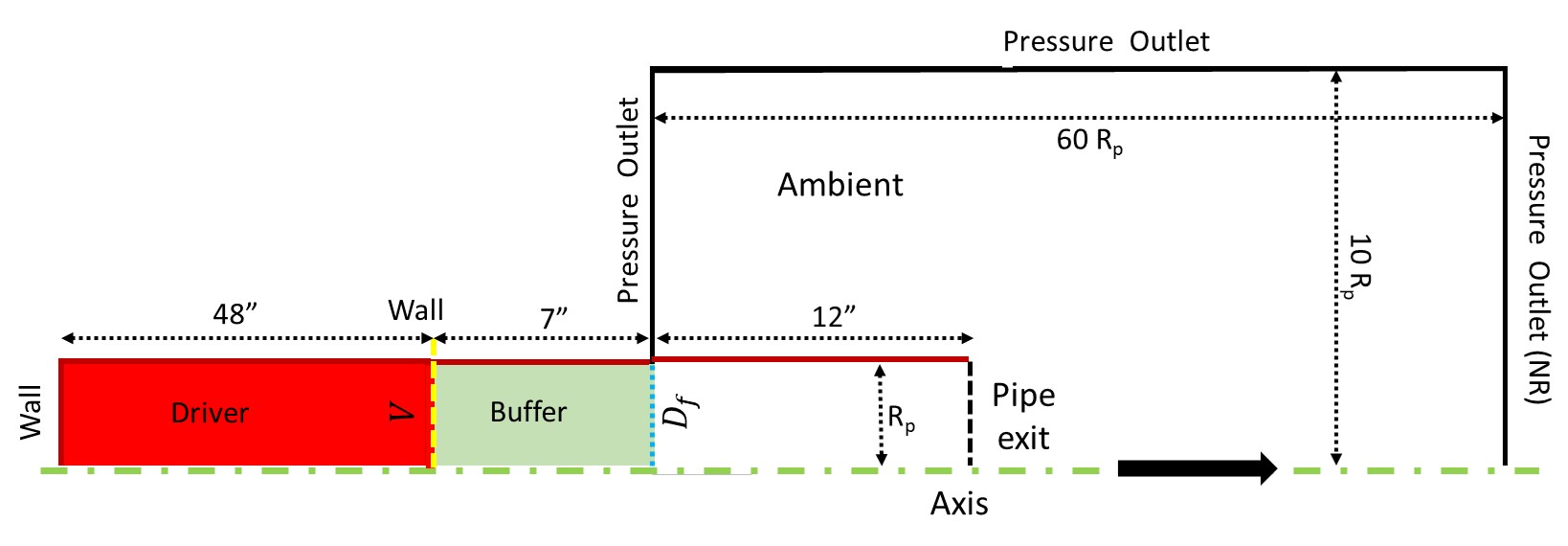}
  \caption{Single phase}
  \vspace{10 pt}
  \label{fig:SimDomains1}
\end{subfigure}%
\\
\begin{subfigure}{.85\textwidth}
  \centering
  \includegraphics[width=1\linewidth]{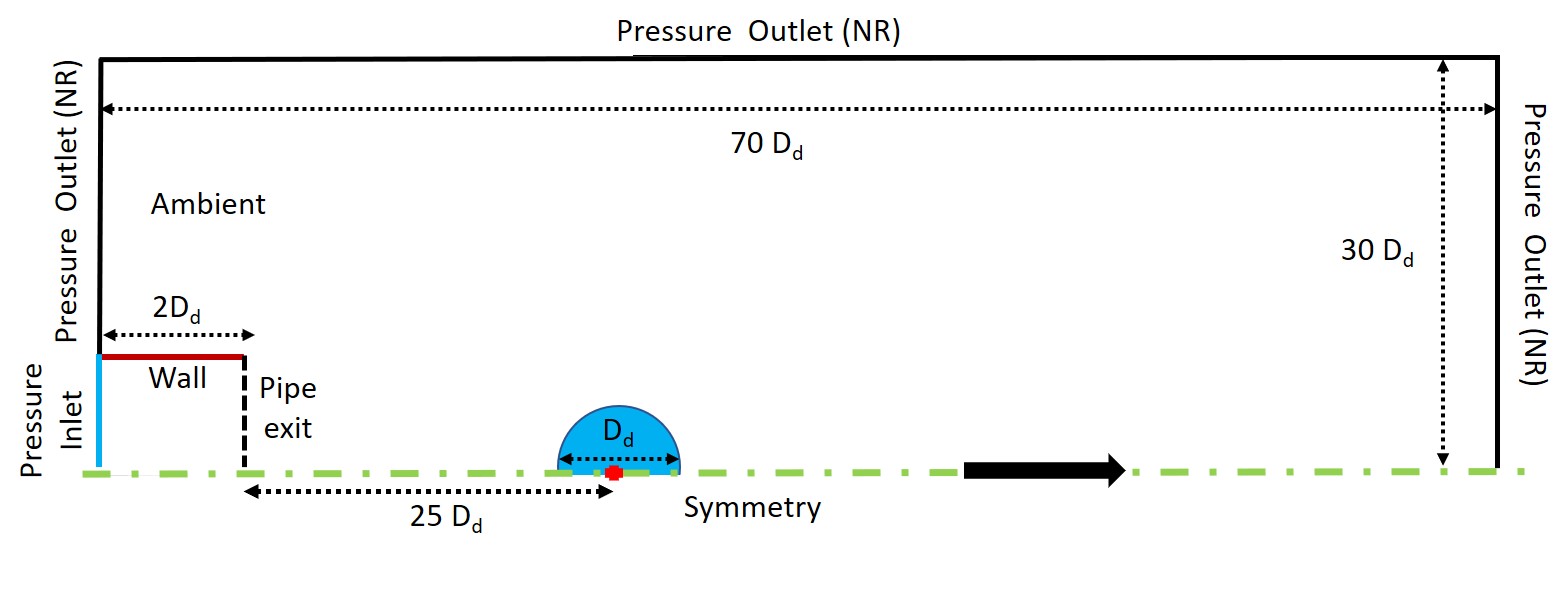}
  \caption{Two-phase}
    \vspace{10 pt}
  \label{fig:SimDomains2}
\end{subfigure}
\caption{Computational domains and boundary conditions for single phase and two-phase axisymmetric simulations.}
\label{fig:SimDomains}
\end{figure}

\subsubsection{Single phase simulation}
Figure~\ref{fig:SimDomains} shows the computational domains and boundary conditions for the single phase and two-phase simulations. In Figure~\ref{fig:SimDomains}, dimensions of the computational domain are scaled by pipe diameter (subscript $p$) or by droplet diameter (subscript $d$). The shock tube is modeled in three sections for the 2D single phase simulations as shown in Fig.~\ref{fig:SimDomains1} consistent with the geometry and dimensions of the experimental setup. Initial conditions are numerically simulated by patching step changes in pressure values in the driver, buffer, and driven sections and treating the diaphragm as a wall. A normal shock front travels in the buffer region towards the diaphragm once the valve is opened, which signifies the start of the simulation. Upon impact by the shock wave, the diaphragm is assumed to undergo an instantaneous rupture, which is numerically simulated by removing the wall boundary condition and redefining the diaphragm as a flow interior interface. In the streamwise direction, a non-reflecting pressure outlet boundary is located sufficiently far from the exit such that it imposes no feedback on the shock tube flow. RANS turbulence model was included for the single phase simulations, which necessitates the non-dimensional distance of cell adjacent to the wall, $Y^+$, to be maintained close to unity. Accordingly, to reduce the overall computational load and time, a biased mesh was employed in the pipe region, with the overall mesh being generated using a block meshing technique in ANSYS- Meshing. Based on a grid convergence study, simulations were conducted on a domain discretized into $\sim$700,000  quadrilateral elements with orthogonal quality of unity.


\subsubsection{Two phase simulation} 
In previous work, simulations have used a 2-D computational domain, which physically describes break-up of a cylindrical column with infinite axial length~\cite{chen08,meng2015numerical}. Another approach used in previous work and also employed in the present work is to use an axisymmetric boundary condition, with the physical system described in cylindrical coordinates~\cite{liou09,chan13,guan18,wadh07}. 
Pressure inlet and non-reflecting pressure outlet boundaries are prescribed as shown in Fig.~\ref{fig:SimDomains2}. A small region of inviscid wall was provided, to separate the pressure inlet and pressure outlet boundaries, which physically represents a region of pipe exit. As shown in the Fig. \ref{fig:SimDomains2}, computational domain dimensions are expressed as a product of droplet diameter, $D_d$. Similar to the single phase flow, block meshing was employed, wherein a region of  4$D_d$ length and 2$D_d$ height around the droplet was uniformly meshed with cell density of 100 cells per diameter, similar to the grid density reported in the work of Stefnitsis \cite{stef21} and Meng et. al \cite{meng2018numerical}, while progressively coarsening the mesh approaching the pressure outlet boundaries.The results reported in the present work are based on the simulations conducted on a computational mesh with $\sim$350,000 orthogonal elements.



\subsection{Numerical methods}\label{subsec32}
\begin{table}[h]
\caption {Numerical models and discretization.}
\centering
\small
\tabcolsep=0.11cm
\begin{tabular}{| p{6cm} | p{6cm} |} 
\hline
   \textbf{Numerical aspect} &\textbf{Solution method} \\ \hline
  PV coupling &	SIMPLEC\\\hline
  Turbulence & SST k-$\omega$ (single phase)\\\hline
  Gradient &	Green Gauss Node based\\\hline
  Pressure &	PRESTO!\\\hline
  Momentum, Energy (convective) & Second order upwind\\\hline
  Multiphase & VOF (Explicit Geo Reconstruct -PLIC)\\\hline
       \hline  
\end{tabular}
\label{Tab:NumModels} 
\end{table}

Table~\ref{Tab:NumModels} summarizes the various numerical models and discretization techniques employed in the single- and two-phase simulations. As mentioned before, single phase simulation included SST k-$\omega$ turbulence model to effectively capture the extent of exit recirculation , shock positions and its effect on pipe exit velocity. For two phase simulations, effects of liquid (droplet) viscosity and surface tension, which were neglected in previous works~\cite{chen08,meng2015numerical,meng2018numerical}, are included. Breakup pattern and time scale can be affected by the viscosity of the droplet \cite{pilch1987use} and its effect can be significant for higher droplet Ohnesorge (Oh) number, implying non trivial viscous effects over inertial and surface tension forces. The Volume Of Fluid (VOF)~\cite{hirt1981volume} method is used to track the droplet-gas interface, given its effectiveness and accuracy in capturing liquid breakup in complex flowfields~\cite{mirjalili2017interface}. No vaporization or turbulence models are considered in the two phase simulations, as the transient variation of velocity and pressure at pipe exit is imposed as boundary condition, based on the single phase simulations. Also, the initial droplet instability and deformation should be unaffected by the turbulence effects, as it is reported to be in good agreement with predictions from previous linear instability studies \cite{sharm21}. A two phase numerical set up, similar to the present work, employed in the work of Stefanitsis et. al \cite{stef21} was shown to have a good agreement with simulation results of Meng et. al \cite{meng2018numerical}.

\section{Results and Discussion}\label{sec4}
\subsection{Shock tube x-t diagram}\label{subsec41}
\begin{figure}[h!]
\centering
\includegraphics[scale=0.4]{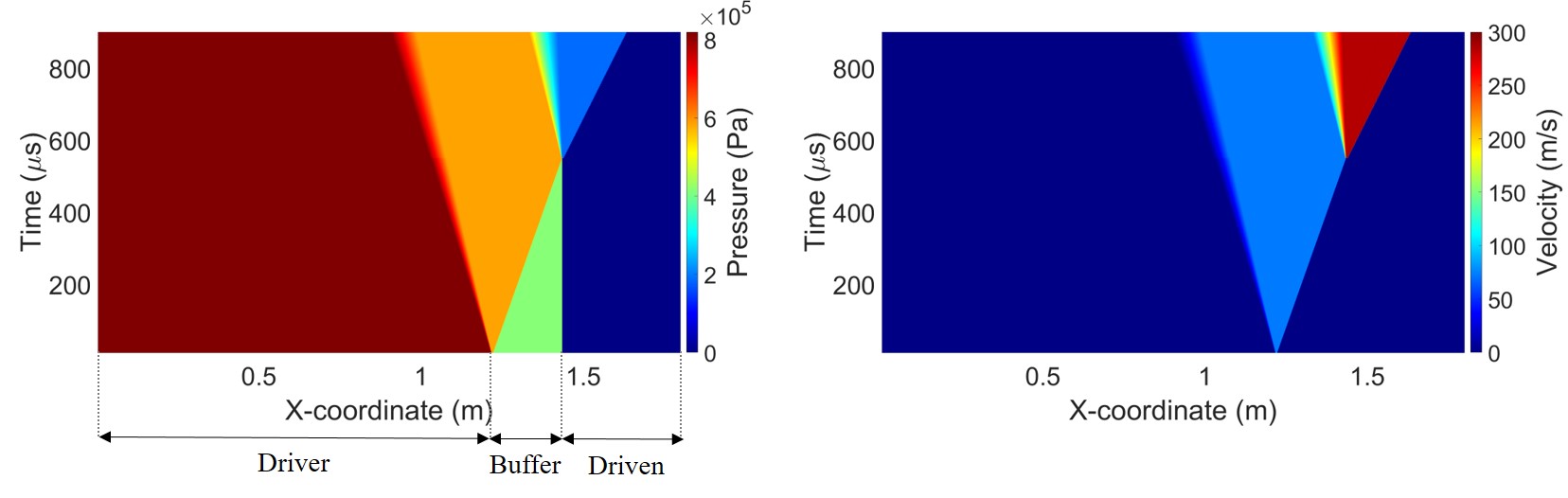}
\caption{x-t diagrams constructed using simulation results.}\label{fig:XTdiagrams}
\end{figure}
Figure~\ref{fig:XTdiagrams} shows x-t diagrams constructed using pressure and velocity contours. As seen from the figure, upon opening of the solenoid valve, a shock wave generates, propagating into the buffer zone, subsequently increasing the pressure in the buffer zone. Upon reaching the driven section interface and following rupture of the diaphragm, which is assumed to occur instantaneously, a shock wave generates propagating into the driven section. The x-t diagrams were constructed using 1D simulations. The primary goal of the x-t diagrams were to determine the influence of changing the intial buffer zone pressure on the strength and speed of the shock wave generated in the driven section. No significant effect was seen to be produced by variation in buffer zone pressure and it was fixed at a constant value of 60 psig for all cases. 
\subsection{Shock tube exit flowfield}
\begin{figure}[h!]
\centering
\includegraphics[scale=0.4]{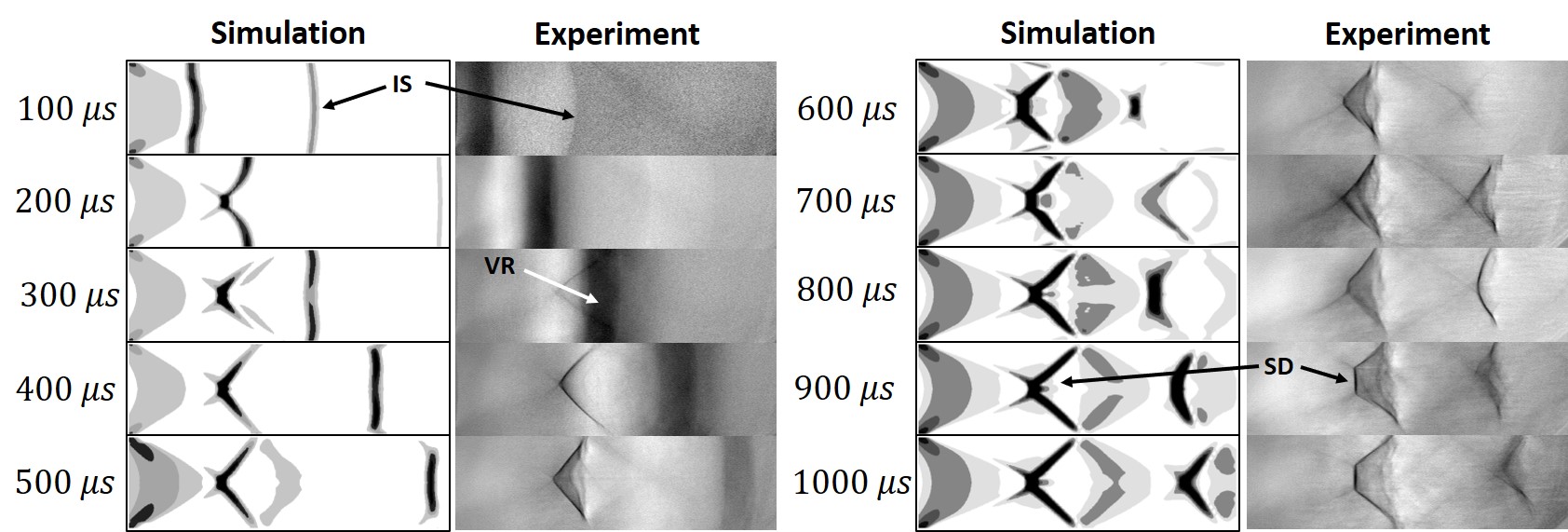}
\caption{Schlieren images from experiment and simulation showing the gas dynamics at the exit of the shock tube.}\label{fig:ExitSchlieren}
\end{figure}
Figure~\ref{fig:ExitSchlieren} shows a time sequence of Schlieren images illustrating gas dynamic features observed at the exit of the shock tube. Experimental images were acquired using the high-speed camera and the Schlieren setup described earlier. Simulation images were obtained from the density field by calculating $\lvert \nabla \rho_x \rvert $~\cite{kawai2010large}. Several interesting features can be noted during the transient evolution of the flow-field into a quasi-steady state at a time point of around 1 ms. A curved incident shock front (IS) emerges from the open end of the tube and propagates in the downstream direction. This can be most clearly observed at t=100 $\mu$s and faintly at t=200 $\mu$s. The speed of this propagating shock front is estimated to be $\sim$ 354 m/s from the experiments and $\sim$ 376 m/s from the simulations giving a Mach number of $\sim$1.08. Following the shock front, a dark region of width $\sim$ 8--10 mm is observed in the experiments, propagating in the downstream direction. This region corresponds to a vortex ring (VR) with an accompanying recirculation zone, which has been observed in previous studies involving open-ended shock tubes~{\cite{elder1952experimental,murugan2018comparative,haselbacher2007open}}. 

\begin{figure}[h!]
\centering
\includegraphics[scale=0.35]{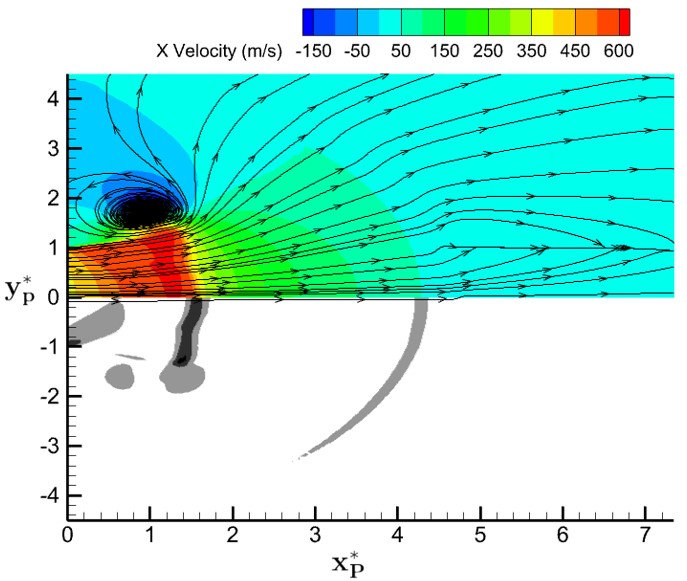}
\caption{Simulation result showing contours of streamwise velocity (top half) and simulated Schlieren (bottom half) at 100 $\mu$s.}\label{fig:100mus}
\end{figure}
Figure~\ref{fig:100mus} shows a simulation result at a time instant of 100$\mu$s where the top half of the image shows contours of the streamwise velocity overlaid with streamlines, while the bottom half shows the simulated Schlieren image obtained from the density field. $x_P^*$ and $y_P^*$ in Fig.~\ref{fig:100mus} are the streamwise and vertical distances non-dimensionalized by the pipe diameter, with $y^*$=0 marking the location of the pipe exit. The recirculation region associated with the vortex ring can be clearly noted in the velocity field contour. Furthermore, the influence of the shock from the Schlieren image on the velocity field can be clearly observed as a step change in velocity behind the shock front. After the initial transients accompanied by the propagating shock front and vortex ring, the flowfield is seen to stabilize by a time duration of around 1 ms. At this time, shock diamonds (SD), characteristic of underexpanded supersonic jets are seen at the exit accompanied by a Mach disk proceeding downstream from the tube exit. It is to be noted that the droplet breakup process as will be discussed in subsequent sections occurs in a shorter time scale, $\mathcal{O}(500 \mu s)$. This implies that the droplet breakup takes place in a transient, developing flowfield. In this sense, the gas flow encountered by the droplet leading to its breakup is different from the uniform gas velocity generated behind a propagating shock front in a closed shock tube.

\subsection{Breakup of single- and two-phase droplets}\label{subsec42}
\begin{figure}[h!]
\centering
\includegraphics[scale=0.45]{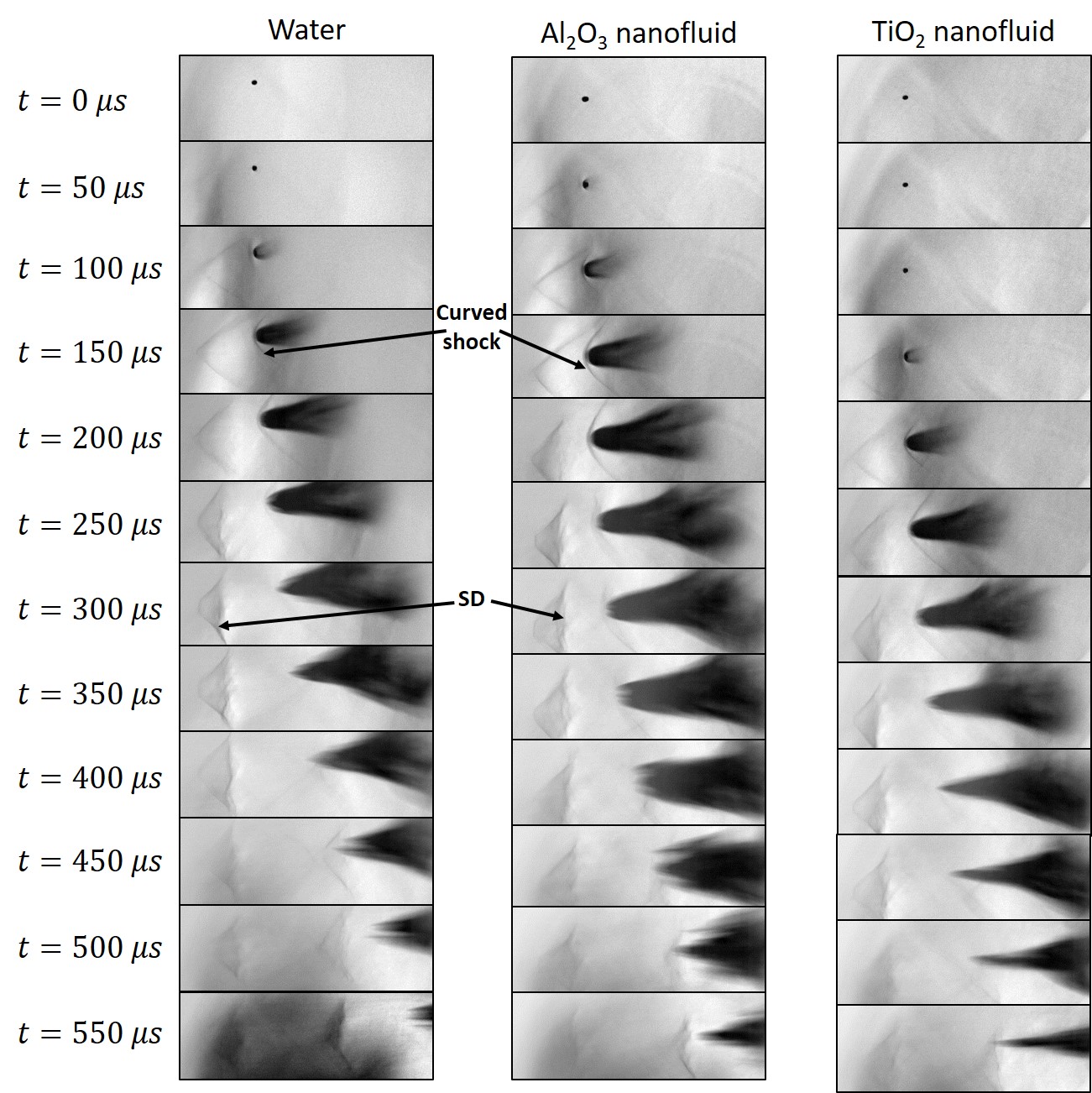}
\caption{Schlieren images showing the sequence of droplet breakup for the three fluids.}\label{fig:SchlierenBreakup}
\end{figure}
\begin{figure}[h!]
\centering
\includegraphics[scale=0.4]{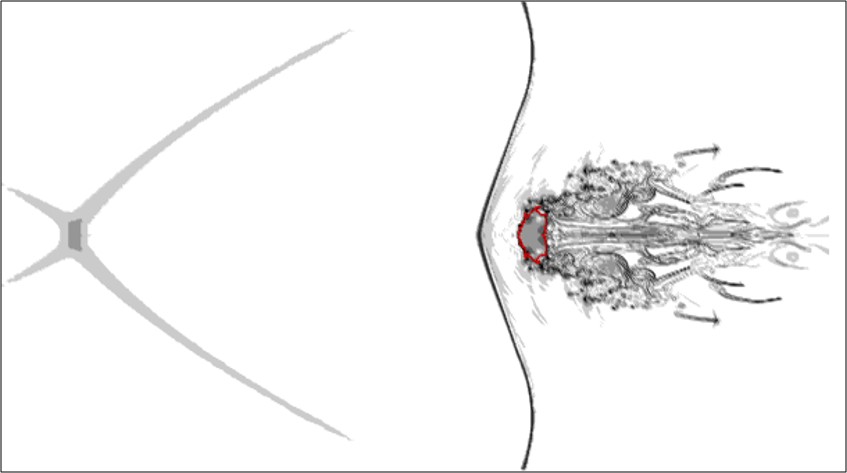}
\caption{Schlieren image from simulations for the case of water droplet at 150 $\mu$s. Droplet edge is marked by a red boundary.}\label{fig:SimSchlieren}
\end{figure}

Figure~\ref{fig:SchlierenBreakup} shows a time sequence of Schlieren images illustrating the droplet breakup process for water and the two nanofluids with properties and at conditions as listed in Table~\ref{Tab:FluidProps}. Start time for the images shown in Fig.~\ref{fig:SchlierenBreakup} is defined corresponding to the instant the droplet shows an initial deformation due to the passage of the curved shock front over it. The shock front and the vortex ring while present for the cases shown in Fig.~\ref{fig:SchlierenBreakup} could not be visualized as clearly as in Fig.~\ref{fig:ExitSchlieren}. This was likely due to the presence of the droplet and ensuing adjustment in contrast of the camera image. However, the phenomena of interest in Fig.~\ref{fig:SchlierenBreakup} is the breakup process of the droplet and the concurrent gas dynamics in the flowfield. In all three cases presented in Fig.~\ref{fig:SchlierenBreakup}, a curved, detached shock front can be observed to be formed at the leading edge of the droplet by about 150 $\mu$s. This shock front is also captured by the simulations as shown in Fig.~\ref{fig:SimSchlieren} along with the overexpanded jet structure further upstream. A turbulent wake is also noticed downstream of the deformed droplet marked by a red boundary in Fig.~\ref{fig:SimSchlieren}. The shock front just upstream of the droplet begins to dissipate by about 250-300 $\mu$s as the droplet breakup proceeds resulting in a mist of droplet mass being formed in the downstream direction. Shock diamonds (SD) observed in the exit flowfield Schlieren images shown in Fig.~\ref{fig:ExitSchlieren} are also observed in the images shown in Fig.~\ref{fig:SchlierenBreakup}. 

\begin{figure}[h!]
\centering
\includegraphics[scale=0.45]{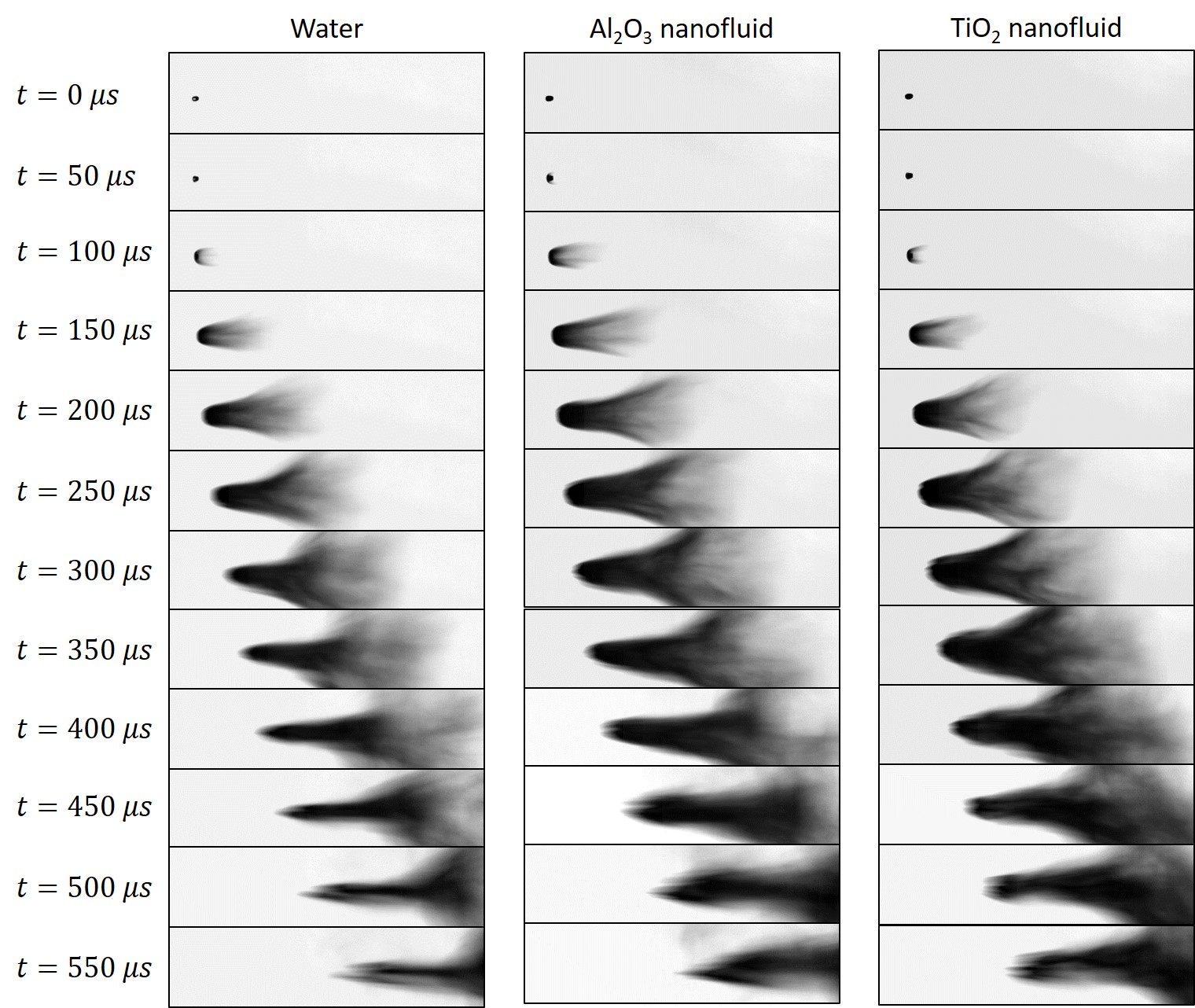}
\caption{High-speed imaging of the droplet breakup process for the three fluids.}\label{fig:DirectBreakup}
\end{figure}
Figure~\ref{fig:DirectBreakup} shows images of droplet breakup for the three fluids as obtained using the direct imaging technique described in Section 2.2. The key difference from the Schlieren images shown in Fig.~\ref{fig:SchlierenBreakup} is the absence of any gas dynamic structures in the field of view. However, a slightly larger field of view could be achieved given the size of the LED display. Further, a clearer contrast between the droplet (including the mist generated by breakup) and the background could be achieved, which made the images from Fig.~\ref{fig:DirectBreakup} more suitable for post-processing to study the dynamics of droplet breakup. Phenomenology of droplet breakup in high speed flow at high We has been extensively analyzed and categorized in previous work particularly those by Ranger~\cite{nicholls1969aerodynamic} (shear stripping or boundary layer stripping), Liu~\cite{liu1997analysis} (sheet thinning), and Theofanous~\cite{theofanous2008physics} (SIE). Simulation results from Meng~\cite{meng2016numerical} provide a very detailed description of the breakup process characterized by: droplet deformation into a ``muffin" shape, generation of a liquid sheet from the droplet periphery, creation of shear layers on either side of the liquid sheet subjecting it to Kelvin-Helmholtz (KH) instabilities, and formation of a non-axisymmetric chaotic wake characterized by vortex shedding, recirculation, and instabilities. The current experimental images shown in Fig.~\ref{fig:SchlierenBreakup} and Fig.~\ref{fig:DirectBreakup} lack sufficient resolution to clearly analyze all the small-scale phenomena reported in previous work. A flattening of the droplet and generation of a ``muffin"-like shape can be observed for all three fluids at the early time (50-100 $\mu$s). 
\begin{figure}[h!]
\centering
\includegraphics[scale=0.6]{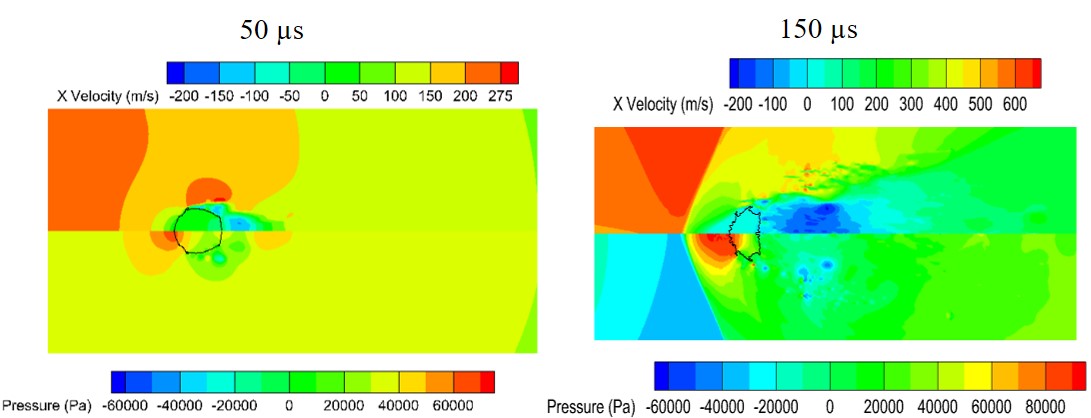}
\caption{Contour plots from simulations for streamwise velocity (\textit{Top}) and pressure (\textit{Bottom}) for the water droplet at 50 and 150 $\mu$s. Coherent edge of the droplet is marked by a black boundary.}\label{fig:SimContours}
\end{figure}

\begin{figure}[h!]
\centering
\includegraphics[scale=0.4]{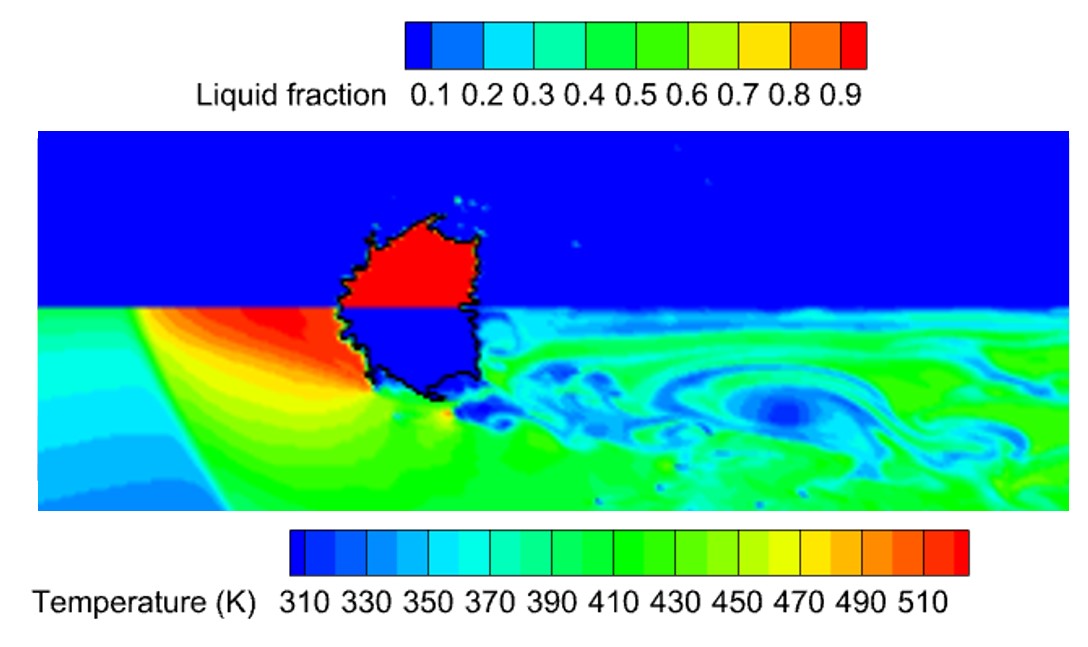}
\caption{Contour plots from simulations for liquid fraction (\textit{Top}) and temperature (\textit{Bottom}) for the water droplet at 150 $\mu$s. Coherent edge of the droplet is marked by a black boundary.}\label{fig:SimContours2}
\end{figure}

This deformation can be observed more clearly in contour plots from simulation results presented in Fig.~\ref{fig:SimContours}, where the coherent edge of the droplet is marked by a black boundary. As in previous work by Meng~\cite{meng2016numerical}, droplet deformation is primarily attributed to the non-uniform pressure distribution resulting in high pressures at the forward and rear stagnation points. This is consistent with the pressure distribution observed in Fig.~\ref{fig:SimContours}. The velocity contours in Fig.~\ref{fig:SimContours} show a corresponding increase in the gas velocity at the lateral edges of the droplet. This further results in a recirculation region referred to by Meng~\cite{meng2016numerical} as the ``equatorial recirculation region", which is responsible for drawing out a liquid sheet from the droplet periphery. The formation of a liquid sheet from the droplet periphery can also be observed in Fig.~\ref{fig:DirectBreakup} at $\sim$ 100 $\mu$s for all three cases. At later times ($>$100$\mu$s), in Fig.~\ref{fig:DirectBreakup}, the liquid sheet generated at the periphery likely envelops the entire droplet and appears as a uniformly dark region with gradual transition to lighter shades in the downstream region populated by droplet mist generated from the breakup. The extended mist formed in the tail of the droplet grows in length and width as time progresses. Within the limits of resolution of the experimental images shown in Fig.~\ref{fig:DirectBreakup}, no significant differences can be observed in the phenomenology of the droplet breakup between the case of water and the two nanofluids. At times instances $\sim$100--150 $\mu$s, the breakup process appears to be most similar to the description by Liu~\cite{liu1997analysis} involving long, thin streamwise ligaments in the droplet wake. This explanation also mentioned by Wang~\cite{wang2020effect} for droplet breakup at high Mach number cites the formation of streamwise ligaments connected by thin films in the droplet wake. Thin films undergo atomization, while the ligaments are stretched by viscous shear and themselves break up in to fragments. Initial stages of  stretching of the droplet film can be better observed in the simulation results presented in Figure~\ref{fig:SimContours2}, which shows contours of liquid fraction as well as gas temperature. The simulations being axisymmetric cannot provide an accurate description of the secondary atomization process. Hence the simulations are primarily used to study the droplet deformation at early stages as well as the gas flow field responsible for the same.

Even as the droplet undergoes secondary atomization, the droplet mass continuously moves in the downstream direction. This motion along with the deformation of the droplet as well as its breakup as characterized by the area of the droplet mist are studied next by post-processing the images shown in Fig.~\ref{fig:DirectBreakup}. 



\subsection{Droplet breakup dynamics}\label{subsec51}
\begin{figure}[h]
\centering
\includegraphics[scale=0.5]{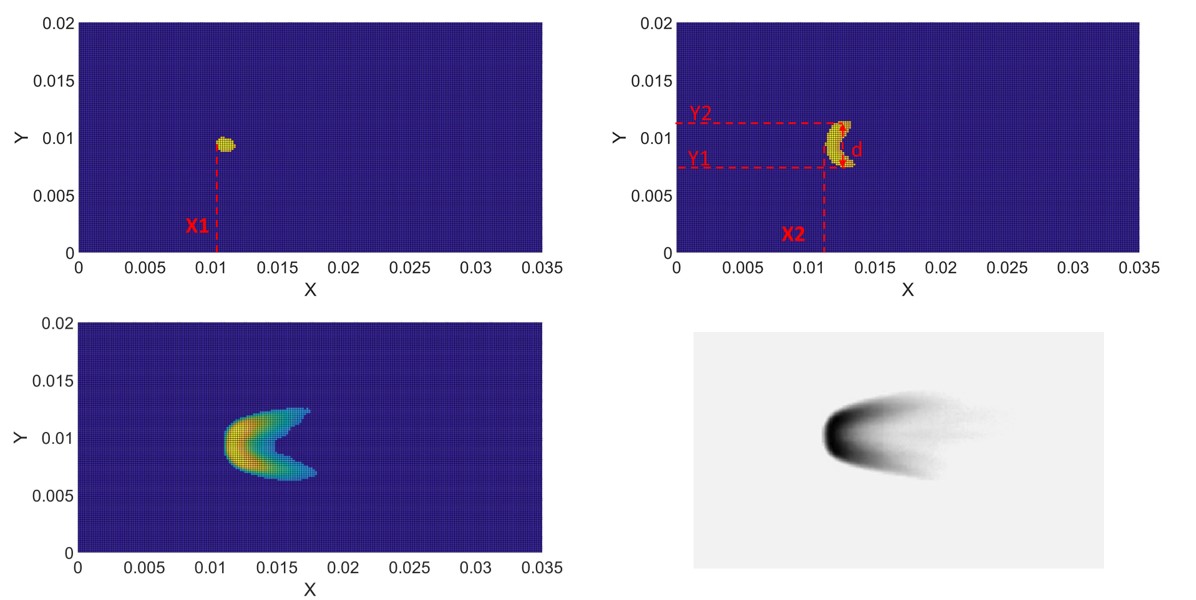}
\caption{Post-processing calculations to find $x^*$, $u^*$, and $d^*$ (\textit{Top}); $A_s^*$ extracted from corresponding experimental image (\textit{Bottom}).}\label{fig:DynaCalc}
\end{figure}

\begin{figure}[h!]
\centering
\includegraphics[scale=0.55]{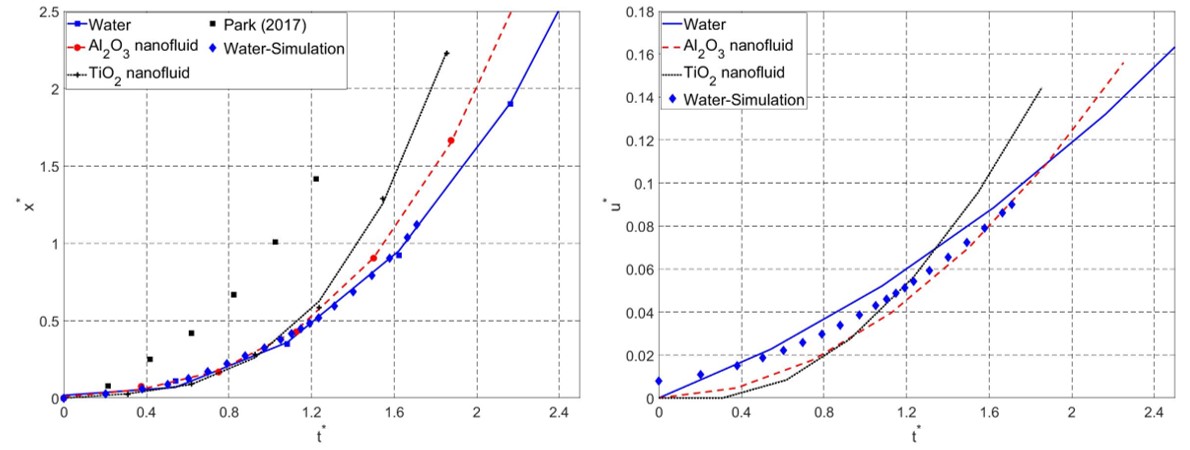}
\caption{Non-dimensional distance and velocity as a function of non-dimensional time for the droplet breakup process for the three fluids.}\label{fig:x*u*}
\end{figure}

\begin{figure}[h!]
\centering
\includegraphics[scale=0.55]{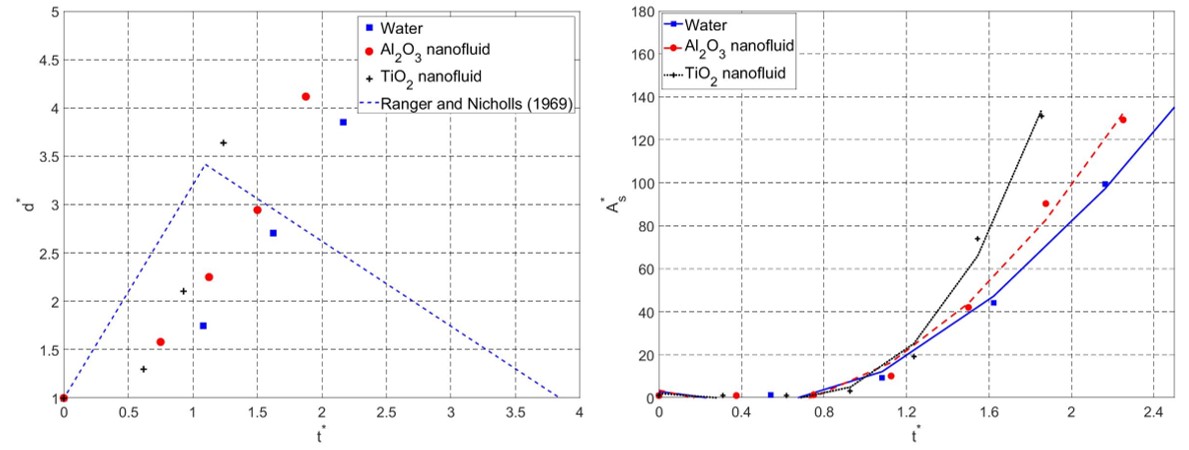}
\caption{Non-dimensional lateral displacement and area of liquid mist as a function of non-dimensional time for the droplet breakup process for the three fluids.}\label{fig:d*A*}
\end{figure}

Figure~\ref{fig:DynaCalc} shows the approach used to estimate various non-dimensional parameters by post-processing images from the experiments as shown in Fig.~\ref{fig:DirectBreakup}. Non-dimensional parameters relevant to droplet breakup dynamics include non-dimensional distance ($x^*=x/d_0$), velocity ($u^*=u/u_g$), time ($t^*=tu_g/d_0\cdot \sqrt{\rho_g/\rho_l}$), lateral displacement ($d^*=d/d_0$) and area of liquid mist ($A_s^*=A_s/A_{s0}$). $x$ and $u$ are the displacement and velocity of a point defined as the leading edge of the droplet, $t$ represents actual time, $d_0$ is the initial length of the major axis of the elliptical droplet, $u_g$ and $\rho_g$ are the gas velocity and density at the pipe exit, $\rho_l$ is the density of the liquid, $d$ is lateral displacement, and $A_s$ is the area of the mist obtained by calculating the weighted sum of pixel area with a minimum grayscale threshold kept constant for all cases. Experimental results shown in Fig.~\ref{fig:x*u*} and Fig.~\ref{fig:d*A*} present an average over three tests.

Figure~\ref{fig:x*u*} shows $x^*$ and $u^*$ for all three fluids with data extracted from the images shown in Fig.~\ref{fig:DirectBreakup} up to a time instant of $\sim$300 $\mu$s. For reference, $x^*$ is also shown as a function of $t^*$ for experimental data obtained by Park~\cite{park17} for breakup of water droplets in a conventional shock tube with a shock Mach number of 1.40. Results for water from the current work follow a similar trend as data by Park~\cite{park17}, however differences in slope arise from the different operating conditions and use of a conventional shock tube in Park's work. Differences are also noted between the behavior of the three fluids tested in the present work. The displacement of the droplet mass is seen to be similar for all three fluids at early times ($t^*<\sim$1.2). However, at later times, a clear increase in displacement of the leading edge of the droplet mass is observed going from water to \ce{Al2O3} nanofluid to \ce{TiO2} nanofluid. With regard to $u^*$, nanofluid droplets are observed to have a smaller non-dimensional acceleration at initial times. However, at later times, $u^*$ for nanofluids gradually exceeds corresponding values for the water droplet, with \ce{TiO2} nanofluid acquiring highest values of $u^*$. The velocity $u^*$ is calculated by fitting a polynomial to $x$, taking its time derivative, and non-dimensionalizing it by $u_g$.

Droplet deformation and breakup under the influence of impulsive flow generated by the open-ended shock tube is a continuous process and proceeds beyond the time duration illustrated in Fig.~\ref{fig:SchlierenBreakup} and Fig.~\ref{fig:DirectBreakup}. However, deformation and breakup during the initial phase of droplet interaction with flow, while the droplet mass is within the field of view of the camera, can be characterized using $d^*$ and $A_s^*$. Figure~\ref{fig:d*A*} shows $d^*$ and $A_s^*$ plotted as a function of $t^*$ for the three fluids. The solid line drawn in Fig.~\ref{fig:d*A*} corresponds to two-part linear approximation proposed by Nicholls~\cite{nicholls1969aerodynamic}. The first part showing an increasing $d^*$ corresponds to the flattening of the droplet resulting in lateral elongation, followed by decrease $d^*$ as a result of fluid mass removal through droplet breakup. Results plotted in Fig.~\ref{fig:d*A*} show that all three fluids in general follow the first part of the trend line suggested by Nicholls~\cite{nicholls1969aerodynamic}. However, given insufficient images within the field of view, the second part of the trend line suggested by Nicholls~\cite{nicholls1969aerodynamic} cannot be conclusively verified. $A_s^*$ shows an increasing trend with $t^*$ as expected. However, in comparing the three fluids, $A_s^*$ for \ce{TiO2} nanofluid is seen to increase faster than that for \ce{Al2O3} nanofluid and water. 

Trends observed in Fig.~\ref{fig:x*u*} and Fig.~\ref{fig:d*A*} for $x^*$, $u^*$, and $A_s^*$ plotted as a function of $t^*$ can be analyzed considering the action of the gas flow in accelerating the droplets as well as causing deformation and mass loss through shearing/stripping mechanisms. 
For an identical force generated by the gas flow field, the denser droplet should experience a slower non-dimensional acceleration, resulting in a smaller initial velocity for the nanofluid droplets. Further, as discussed by Shen~\cite{shen2019viscous}, for a fluid with higher viscosity resulting in a droplet with higher Oh, droplet breakup time is delayed. This occurs due to more of the inertial force of the flow being consumed by viscous dissipation for the higher viscosity fluid. Concurrently, for the case with higher Oh, the droplet mass will stay coherent for a longer duration. Results for $d^*$ shown in Fig~\ref{fig:d*A*} indicate the lateral deformation for nanofluid droplets to be larger than that for water. This likely indicates that the nanofluid droplets tend to undergo larger lateral deformation and stay coherent for a longer duration than water droplets, which start to undergo breakup earlier due to lower Oh. The deformed yet coherent nanofluid droplet can thus experience a larger acceleration compared to water droplet at later times causing the highest values of $x^*$ and $u^*$ to occur for \ce{TiO2} nanofluid as seen in Fig.~\ref{fig:x*u*}.

\begin{figure}[h!]
\centering
\includegraphics[scale=0.4]{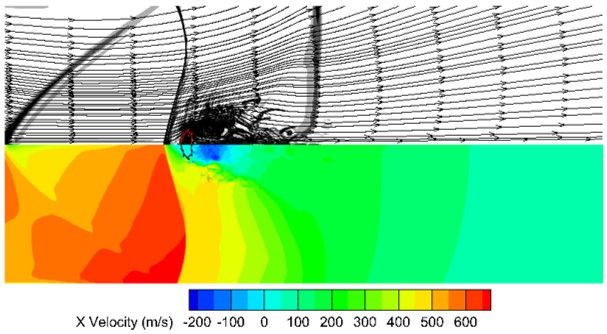}
\caption{Two-phase simulation results for a water droplet showing numerical Schlieren overlaid with streamlines (\textit{Top}) and streamwise velocity contours (\textit{Bottom}) at 150 $\mu$s.}\label{fig:SimContours3}
\end{figure}

Further, as suggested by Wang~\cite{wang2020effect}, a higher Oh would correlate with generation of larger sheets and ligaments. The gas flowfield in the wake of the droplet undergoing breakup is dissimilar from a similar flowfield established with the use of a conventional shock tube. Figure~\ref{fig:SimContours3} shows simulation results highlighting the position the detached curved shock and the droplet undergoing breakup using a numerical Schlieren along with streamlines indicating the direction of gas flow. As can be noticed from the figure, the unconstrained nature of the gas flow generated from the open ended shock tube causes the flow to expand outwards unlike the streamwise-aligned flow generated in a conventional shock tube~\cite{wagner2016pulse}. This flow profile is the likely cause for the lateral spread of the liquid mist observed downstream of the main droplet mass. This results in an increase in $A_s^*$ for all cases. However $A_s^*$ is highest for \ce{TiO2} nanofluid due to its tendency to break up into larger fragments experiencing higher acceleration. 

\section{Conclusions}\label{sec7}
Aerobreakup of single- and two-phase liquid droplets are studied using experiments and numerical simulations. Experiments employ high-speed imaging of the droplet breakup process caused by impulsively generated high speed flow of air using an open ended shock tube. Axisymmetric numerical simulations performed only for the single-phase liquid droplets use a Volume-Of-Fluid technique to capture droplet surface deformation and gas dynamics in the flowfield. High-speed Schlieren imaging captures unique features in the transient flowfield at the exit of the open ended shock tube including a leading edge incident shock followed by a recirculation zone and shock diamonds similar to those in underexpanded jets. These features, which are also observed in the simulation results generate a significantly different flowfield for droplet aerobreakup as compared to conventional closed shock tubes. Droplet breakup for water, \ce{Al2O3} nanofluid, and \ce{TiO2} nanofluid is investigated at identical flow conditions. Imaging results from experiments as well as simulation results show the droplet breakup process to be accompanied by the formation of a curved leading edge shock front, flattening of the droplet, generation of a liquid sheet and its subsequent atomization into a fine mist under the influence of shear forces. No significant qualitative differences are observed in the phenomenology of droplet breakup for two-phase fluids. However, the mist in the wake of the coherent droplet mass for all cases is observed to grow laterally in contrast with similar observations made in closed shock tubes. This is likely caused by outwardly expanding flow generated at the exit of an open ended shock tube unlike the streamwise flow behind a propagating shock wave in a conventional shock tube. The motion of the droplet mass as it undergoes breakup is studied by considering its non-dimensional position, velocity, deformation, and area of spread as a function of non-dimensional time. A key finding is that, past an initial period, the \ce{TiO2} nanofluid appears to show the highest velocity and area of spread followed by \ce{Al2O3} nanofluid and water. This mirrors the ordering of the Oh number of the three cases, which itself is influenced by fluid viscosity for other factors staying the same. \ce{TiO2} nanofluid has the highest viscosity followed by \ce{Al2O3} nanofluid and water. High Oh number is associated with delay in droplet aerobreakup as well as generation of larger fragments. The delay in breakup is seen to manifest in larger lateral deformation of the nanofluid droplets as compared to that of water, and subsequently higher velocities with the maximum value being for \ce{TiO2} nanofluid. Higher viscosity droplets also break up into larger fragments, which combined with their higher acceleration results in a larger area of spread. It can be thus concluded that aerobreakup of nanofluids proceeds similar to those of pure fluids with higher effective viscosity and hence Oh number. An open ended shock tube generates droplet aerobreakup similar to that in a conventional shock tube except for the dispersion of fine mist in the droplet wake, which increases in the lateral direction due to the particular gas dynamics of the flowfield for the open ended shock tube.
\backmatter





\bmhead{Acknowledgments}

Partial funding for this work was provided by a LaSPACE graduate student research assistance award. The contribution of Ms. Lailah Collins and Ms. Wanjun Dang towards this work are acknowledged.



 \section*{Declarations}


 \begin{itemize}
 \item Funding: Partial funding for this work was provided by a LaSPACE graduate student research assistance (GSRA) award number AWD-AM200738.
 \item Conflict of interest/Competing interests: All authors certify that they have no affiliations with or involvement in any organization or entity with any financial interest or non-financial interest in the subject matter or materials discussed in this manuscript.
 \item Availability of data and materials: The datasets generated during and/or analysed during the current study are available from the corresponding author on reasonable request.
 \end{itemize}

\bibliography{sn-bibliography}


\end{document}